# ACOUSTICAL PROPERTIES OF SUPERFLUID HELIUM IN CONFINED GEOMETRY.

**Sh. E. Kekutia and N. D. Chkhaidze.**

Institute of Cybernatics, 5 Sandro Euli, Tbilisi, 0186, Georgia

E-mail: kekuka@yahoo.com; nikolozi1951@yahoo.com

**CONTENTS**

**Abstract**

The problem studied in this paper is to obtain the equations describing sound propagation in a consolidated porous medium filled with superfluid $He^4$, determine the elastic coefficients, appearing in the equations, in terms of physically measurable quantities, and calculate the propagation velocities of transverse and longitudinal waves at high and low oscillating frequencies. In general, the obtained equations describe all volume modes that can propagate in a porous medium saturated with superfluid $He^4$ for any values of the porosity and frequencies $(0 \leq f \leq 1;\ 1 \leq a_\infty < \infty)$.

The derived equations are applied to the most important particular case when the normal component of superfluid helium is locked inside a highly porous media (aerogel, Im-helium sample) by viscous forces. For this case the velocities of two longitudinal sound modes and transverse mode are calculated from the derived equations. There are established the coupling between temperature and pressure oscillations in these fast and slow modes.

# 1. INTRODUCTION

The investigation of sound propagation in nanostructured materials filled with superfluid helium is a rapidly developing research field. This is not only because of technological importance studied system, but also because it provides opportunities to examine the effects of complex, disordered pore structures on the properties of the fluids confined in porous media.

There are some basic approaches to studying sound propagation in the physical system *HeII* in a porous medium. In the first one the superfluid helium fills the porous solid body, which can be taken as absolutely rigid, i.e. the solid body does not participate in the oscillating motion of the liquid. This direction of investigation first appeared in 1948 in Ref. 1, where attention was first called to the fact that the character of sound propagation in superfluid helium can be strongly altered by retarding the motion of the normal component in the helium. Fourth sound propagates in this case [2]. The first experimental measurements [3,4] of the propagation velocity of fourth sound were described well by a formula derived from two-fluid linearized system of equations of hydrodynamics. The velocity of the normal component of the fluid was set equal to zero and the equation describing the law of conservation of momentum was eliminated. These experiments were performed on a complicated system of branched capillaries. It is very difficult to investigate such a system theoretically. A quite complete theory of sound propagation in a simple porous medium was constructed in Ref.5 and 6, where sound propagation was studied in isolated capillaries oriented parallel to one another. The experimental results, which agreed well with this theory, are published in Ref. 7 and 8.

The second direction of research is more recent and arose after the first experiments on sound propagation in an easily entrained porous medium – an aerogel [9-11]. We note that aerogel in superfluid has been studied very intensively during last decade [ 9, 10, 12, 13, 14, 15]. Aerogels are a unique class of ultra fine cell size, have low density, and are open-cell foams. Aerogels have continuous porosity and a microstructure composed of interconnected colloidal-like particles or polymeric chains with characteristic diameters of 100 angstroms. The microstructure of aerogels is responsible for their unusual acoustic, mechanical, optical, electrical and thermal properties. These microstructures impart high surface areas to the aerogels, for example, from about 350 $m^2$/g to about 2000 $m^2$/g. There ultra fine cell/pore size minimizes light scattering in the visible spectrum, and thus, aerogels can be prepared as transparent porous solids. Further, the high porosity of aerogels makes them excellent insulators with their thermal conductivity being about 100 times lower than that of the prior known dense matrix foams. Still further, the aerogel skeleton provides for the low sound velocities observed in aerogels. Currently, aerogels of various compositions are known, and these aerogels were generally referred to as inorganic (such as silicon aerogels) and organic (such as carbon aerogels). Such inorganic aerogels, as silica, alumina, or zirconia aerogels, are traditionally made by means of the hydrolysis and condensation of metal alkoxides, such as tetramethoxy silane. Organic aerogels, such as carbon aerogels, are typically made from the sol-gel polymerization of resorcinol or melamine with formaldehyde under alkaline conditions. Therefore it is obvious that, when aerogel is saturated even with pure $HeII$ new phenomena are caused by the presence of aerogel: namely, the coupling between the two sound modes is provided by $\rho^a/\rho$ ($\rho^a$ -is the aerogel density) [11], which enhances the coupling between the two sound modes. Such replacement of coupling between sounds accompanies by change of sound conversion character in impure superfluids and for superfluids in aerogel. For example, the propagation of a heat pulse with the velocity of the first mode in $HeII$ in aerogel has been observed [14].

The biggest success to nanostructured porous materials saturated with superfluid helium is stipulated by development of new techniques for producing impure superfluids, which exhibit unique properties. The introduction of the impurity particles into liquid helium produces impurity-helium (Im-$He$) clusters, which make it possible to create macroscopic Im-$He$ samples consisting of impurity atoms isolated in liquid or solid helium. At first these systems were obtained by injecting atoms and molecules such as $N_e, Ar, Kr, D_2, N_2, H_2O, C_2H_5OH, Ba, Na$ [16, 17, 18, 19]. Superfluid helium confined to aerogel [9, 14], superfluid in Vycor glasses [20] and watergel-a frozen water "lattice" in $HeII$ [21] can also be related to this new class of systems. Usually the macroscopic samples of the Im-$He$ solid phase are built from aggregations of small Im-$He$ clusters. Furthermore, these aggregates form extremely porous solids into which liquid helium can easily penetrate. These porous solids consist of a loosely or tightly connected continuous network of impurities or clusters of impurities each of which is surrounded by one or two layers of solidified helium. Therefore there is a unique opportunity to investigate the properties of superfluid helium in porous nanostructures.

The third direction is based on the works of Biot [22] for classical liquid filling a «partially entrained» porous medium. Our work concerns this more general direction, the first two directions being particular cases of the later. The Biot theory predicts propagation of two different volume longitudinal waves and one shear wave. There waves are solutions of two coupled differential equations which describe the motion of a two-component system. The parameters of the Biot theory are the porosity $f$, the tortuosity $a$, the fluid density $r_f$, and the solid body density $r_{sol}$, the bulk modulus of the fluid $K_f$ and of the solid $K_{sol}$, and the bulk modulus $K_b$ and the shear modulus $N$ of the «dry» sample [22, 24, 25]. It should be noted that if the solid framework is very stiff, i.e. $K_b, N \gg K_f$, the phase velocity of the slow longitudinal wave is $C = \dfrac{V_f}{\sqrt{a_\infty}}$, where $V_f = \sqrt{K_f/\rho_f}$ is the sound velocity in the liquid and $a_\infty$ is the tortuosity at high oscillation frequencies. In this limit the slow wave is also the wave propagating in a liquid, the difference being that because of the tortuosity of the channel it becomes modified. This wave was observed first in ref.26 in a porous medium consisting of sintered water-filled glass beads (consolidated porous medium). The porosity of the samples ranged from 7.5 to 28.3%; the measerements were performed using the ultrasonic load conversion technique.

Biot's theory of wave propagation in a fluid-filled porous medium has been used [27] to study the propagation of sound in superfluid $He^4$ in a porous medium. In Ref. 27 it was assumed that at T<1K, when the density of the normal component of $HeII$ can be set equal to zero, fourth sound in helium corresponds to the slow longitudinal wave arising in the Biot theory. The coefficient n, due to multiple scattering of sound in a porous medium and playing a large role in the description of the measerements of the velocity of sound, is related with structure factor $a_\infty$ in the Biot theory by the relation $n^2 = a_\infty$. The values of the velocity of the corresponding waves in $HeII$ were predicted on the basis of the results obtained from the experimental study of Biot waves in water filling a porous structure. The velocity of the slow longitudinal wave agree well with the velocity of fourth sound.

Landau predicted the existence of second sound in free $HeII$ [28]. Peshkov was the first to observe the propagation of second sound [29]. Singer et al. studied experimentally in the temperature range 1,3-$T_l$ second sound in $HeII$ filling the channels of a porous medium. The porous medium consists of sintered 180-210 $\mu m$ glass particles with porosity ranging from 16 to 33%. Capacitive transducers with a vibrating porous membrane served as emitter and detector of second sound. It was shown that the second sound velocity in a porous medium is lower than the bulk velocity. The parameter n was determined in Ref. 30 from these data.

The ultrasonic properties of unconsolidated (in the ideal case a suspension) and consolidated porous media saturated with water have also been investigated [31]. Only one wave was observed in the first case; a slow and a fast waves propagated in the second case. The ultrasound emission technique based on the concept of mutual conversion and refraction at a liquid-solid interface, was used to excite volume waves in a solid layer of the material. The acoustics of $HeII$ filling a porous medium consisting of spherical glass grains with 38% porosity was analysed. The results obtained in Ref. 31 for an superleak consisting of compressed $Al_2O_3$ powder with 65% porosity and filled with $HeII$ agreed with conjectures made in Ref. 27.

Investigations of transverse waves in a porous solid-liquid system are very informative. The authors of Ref. 32 probed various ceramic materials with superfluid helium. The porosities of the samples were 44, 45, and 92%. The different pore sizes and the wide spectrum of permeabilities made it possible to perform measerements in the high and low-frequency limits of the oscillations of the thermodynamic quantities. Ultrasonic measerements were performed using a phase-sensitive pulsed technique. The results obtained made it possible to determine the structural parameters of the porous medium: the tortuosity, the permeability, and the effective size of pores. In Ref. 32 specific porous media filled with $HeII$ were studied, and Biot's approach must be used in order to investigate simultaneously the transverse and longitudinal waves propagating in such systems.

The experimental and theoretical works presented above indicate that the Biot theory needs to be extended to the case where superfluid helium saturates a porous medium. In a superfluid liquid, in contrast to an ordinary liquid, several waves propagate. In the present paper the propagations of waves is studied in a three-component system porous medium-superfluid liquid in a wide range of porosities. This made it possible to study the properties of the porous medium in greater detail [12, 13]. The problem studied in this paper is to obtain the equations describing sound propagation in a consolidated porous medium filled with superfluid $He^4$, determine the elastic coefficients, appearing in the equations, in terms of physically measurable quantities, and calculate the propagation velocities of transverse and longitudinal waves at high and low vibrational frequencies. In general, the obtained equations describe all volume modes that can propagate in a porous medium saturated with superfluid $He^4$ for any values of the porosity and frequencies $(0 \leq f \leq 1;\ 1 \leq a_\infty < \infty)$.

The derived equations are applied to the most important particular case when the normal component of superfluid helium is locked inside a highly porous media (aerogel, Im-helium sample) by viscous forces. For this case the velocities of two longitudinal sound modes and transverse mode are calculated from the derived equations. There are established the coupling between temperature and pressure oscillations in these fast and slow modes.

## 2. EXPRESSION OF GENERALIZED COEFFICIENTS BY PHYSICALLY MEASURED QUANTITIES.

The elastic properties of a system containing a superfluid helium completely filling the pores were considered in [12, 13], where methods for measurement of generalized elastic coefficients are described with jacketed and unjacketed compressibility tests in the case of a homogeneous and isotropic porous matrix. According to [12, 13] the stress-strain relations are

$$\begin{aligned}
s_x &= 2Ne_x + Ae + Q^S e^S + Q^n e^n \\
s_y &= 2Ne_y + Ae + Q^S e^S + Q^n e^n \\
s_z &= 2Ne_z + Ae + Q^S e^S + Q^n e^n \\
t_x &= Ng_x, t_y = Ng_y, t_z = Ng_z. \\
s' &= Q^S e + R^S e^S + R^{Sn} e^n \\
s'' &= Q^n e + R^n e^n + R^{Sn} e^S
\end{aligned} \qquad (1)$$

where $s_x, s_y, s_z$ and $t_x, t_y, t_z$ are normal and tangential forces acting on an element of the solid surface with the following orientation, $s'$ and $s''$ - forces acting on the liquid part, which correspond to superfluid and normal components of superfluid helium.

Since the present system is a porous structure, we assume that the unit volume is much larger than the pore size. Therefore we determine the displacement vector $\vec{u}$ as the displacement of the solid averaged over a volume element. The average displacement vector $\vec{U}$ of the liquid part of the cube, which determines the fluid flow, can be determined in the same manner.

The average displacement vector of the solid has the components $u_x, u_y, u_z$ and that of the superfluid helium $U^S_x, U^S_y, U^S_z, U^n_x, U^n_y, U^n_z$. The solid strain components are then given by

$$e_x = \frac{\partial u_x}{\partial x}, \; e_y = \frac{\partial u_y}{\partial y}, \; e_z = \frac{\partial u_z}{\partial y}, \; g_x = \frac{\partial u_y}{\partial z} + \frac{\partial u_z}{\partial y}, \; g_y = \frac{\partial u_x}{\partial z} + \frac{\partial u_z}{\partial x}, \; g_z = \frac{\partial u_x}{\partial y} + \frac{\partial u_y}{\partial x} \quad (2)$$

Due to two possible types of motion in $He\,II$ the displacement of superfluid solution $\vec{U}$ breaks down into the sum of two parts

$$\vec{U} = \frac{r^S}{r}\vec{U}^S + \frac{r^n}{r}\vec{U}^n \quad (3)$$

corresponding to displacement of superfluid and normal components. Thus the strain in fluid is defined by the dilatation

$$e = \frac{r^S}{r}\nabla \vec{U}^S + \frac{r^n}{r}\nabla \vec{U}^n \quad (4)$$

Since the superfluid and normal components cannot be physically separated in $HeII$ and it is meaningless to talk about whether individual atoms of the liquid belong to the superfluid or normal components, the following relation should take place:

$$Q^S e^S + Q^n e^n = Qe \quad (5)$$

The coefficient A and N correspond to the well-known Lame coefficient in the theory of elasticity and are positive [24]. The coefficient Q and R are the familiar Biot's coefficients [23]. $R^S(R^n)$ is a measure of the stress arising in the superfluid (normal) component when a unit volume of the system is compressed without compressing the normal (superfluid) component and the porous medium. The coefficient $R^{Sn}$ determines the stresses arising in the superfluid component when the normal component is compressed without compression of the superfluid component and the porous medium, and vice versa [12].

Let us consider Hypothetical experiments which make it possible to determine the generalized elastic coefficients in terms of the measured coefficients $K_s, K_f, K_b$ and N. Such an approach was proposed in Ref. 23 for a normal liquid.

In the first experiment the porous sample was placed in a superfluid liquid to which pressure $p'$ was applied. Under pressure liquid completely penetrates into the pores and the quantities ? and $\mathbf{e}$ are measured. Therefore $K_s$ and $K_f$ can be determined as

$$\frac{1}{K_s} = -\frac{e}{p'}, \quad \frac{1}{K_f} = -\frac{\mathbf{e}}{p'} \tag{6}$$

and according to Eq. (1) we cam write

$$\mathbf{s}_x = \mathbf{s}_y = \mathbf{s}_z = -(1-\mathbf{f})p'$$
$$s' = -\mathbf{f}\frac{\mathbf{r}^s}{\mathbf{r}}p' \tag{7}$$
$$s'' = -\mathbf{f}\frac{\mathbf{r}^n}{\mathbf{r}}p'$$

We shall also use the condition $\mathbf{e}^s = \mathbf{e}^n = \mathbf{e}$. Then the relations (7) can be put into the form:

$$\frac{2}{3}Ne + Ae + Q\mathbf{e} = -(1-\mathbf{f})p'$$
$$\frac{\mathbf{r}^s}{\mathbf{r}}Q\,e + (R^s + R^{sn})\mathbf{e} = -\mathbf{f}\frac{\mathbf{r}^s}{\mathbf{r}}P' \tag{8}$$
$$\frac{\mathbf{r}^n}{\mathbf{r}}Q\,e + (R^n + R^{sn})\mathbf{e} = -\mathbf{f}\frac{\mathbf{r}^n}{\mathbf{r}}P'$$

The last two equations give

$$R^s + R^{sn} = K_f \frac{r^s}{r}\left(f - \frac{Q}{K_f}\right)$$
$$R^n + R^{sn} = K_f \frac{r^n}{r}\left(f - \frac{Q}{K_f}\right) \tag{9}$$

In the second experiment the porous solid sample is enclosed in a thin impermeable shell and is subjected to the pressure $p'$ of the liquid. To ensure a constant pressure inside the liquid must be flow out through a tube into an external reservoir. Thus $e^s = e^n = e$ and $K_b$ is given by the relation

$$\frac{1}{K_b} = -\frac{e}{p'} \tag{10}$$

in this experiment

$$\sigma_x = \sigma_y = \sigma_z = -p'$$
$$s' = s'' = 0 \tag{11}$$

or

$$\left(\frac{2}{3}N + A\right)\frac{1}{K_b} - Q\frac{e}{p'} = 1$$
$$\frac{r^s}{r K_b}Q - (R^s + R^{sn})\frac{e}{p'} = 0 \tag{12}$$
$$\frac{r^n}{r K_b}Q - (R^n + R^{sn})\frac{e}{p'} = 0$$

whence

$$r^n(R^s + R^{sn}) = r^s(R^n + R^{sn}) \tag{13}$$

It is easy to show from Eqs. (8), (9) and (12) that [23]

$$Q = \frac{fK_{sol}\left(1 - f - \frac{K_b}{K_{sol}}\right)}{1 - f + f\frac{K_{sol}}{K_f} - \frac{K_b}{K_{sol}}}$$

$$\frac{2}{3}N+A = K_{sol}\frac{(1-f)\left(1-f-\dfrac{K_b}{K_{sol}}\right)+f\dfrac{K_b}{K_f}}{1-f+f\dfrac{K_{sol}}{K_f}-\dfrac{K_b}{K_{sol}}} \qquad (14)$$

In the third experiment a superleak provides the coupling with the reservour. This leak passes only the superfluid component, so that

$$\begin{aligned}s_x = s_y = s_z &= -(1-f)p' \\ s' &= 0 \\ s'' &= -f p'\end{aligned} \qquad (15)$$

or

$$\begin{aligned}\left(\tfrac{2}{3}N+A\right)e+\frac{r^s}{r}Qe^s+\frac{r^n}{r}Qe^n &= -(1-f)p' \\ \frac{r^s}{r}Q\,e+R^s e^s+R^{sn} e^n &= 0 \\ \frac{r^n}{r}Q\,e+R^n e^n+R^{sn} e^s &= -f p'\end{aligned} \qquad (16)$$

In this case the relation between $e^s$ and $e^n$ can be obtained using the laws of conservation of mass and entropy.
Then

$$e^s - e^n = \frac{C_{He}\,p'}{r^s s^2 T}, \qquad (17)$$

where $C_{He}$ and $s$ are the specific heat and entropy per unit mass of the $He\,II$. Finally, after laborious calculations we find the desirable relations

$$\begin{aligned}R^{sn} &= \frac{r^s r^n}{r^2}R - \frac{f(r^s)^2 T s^2}{r\,C_{He}} \\ R^s &= \frac{(r^s)^2}{r^2}R + \frac{f(r^s)^2 T s^2}{r\,C_{He}} \\ R^n &= \frac{(r^n)^2}{r^2}R + \frac{f(r^s)^2 T s^2}{r\,C_{He}}\end{aligned} \qquad (18)$$

Where the Biot-Willis coefficient R is [23]

$$R = \frac{f^2 K_s}{1 - f + f \frac{K_s}{K_f} - \frac{K_b}{K_s}} \qquad (19)$$

We note once again that the coefficients $K_f$, $K_s$, $K_b$ and $N$ are experimentally measurable quantities.

## 3. EQUATIONS OF MOTION

We shall use LaGrange's formalism to find the equations of motion of the system porous solid-superfluid liquid. To construct the desired equations the kinetic energy of the system must be determined after the generalized coordinates are chosen. We assume that the physical point is a region of size much greater than the pore size but much smaller than the characteristic lengths in the problem (for example, the wavelength when considering wave processes). For the generalized coordinates of the system we choose the nine components of the displacement vectors of the liquid and solid, averaged over the volume of the physical point: $u_x, u_y, u_z, U^s_x, U^s_y, U^s_z, U^n_x, U^n_y, U^n_z$. Then the kinetic energy $T$ per unit volume can be represented as

$$2T = r_{11}\left[\left(\frac{\partial u_x}{\partial t}\right)^2 + \left(\frac{\partial u_y}{\partial t}\right)^2 + \left(\frac{\partial u_z}{\partial t}\right)^2\right] + 2r^s_{12}\left[\frac{\partial u_x}{\partial t}\frac{\partial U^s_x}{\partial t} + \frac{\partial u_y}{\partial t}\frac{\partial U^s_y}{\partial t} + \frac{\partial u_z}{\partial t}\frac{\partial U^s_z}{\partial t}\right] +$$

$$+ 2r^n_{12}\left[\frac{\partial u_x}{\partial t}\frac{\partial U^n_x}{\partial t} + \frac{\partial u_y}{\partial t}\frac{\partial U^n_y}{\partial t} + \frac{\partial u_z}{\partial t}\frac{\partial U^n_z}{\partial t}\right] + 2r^s_{22}\left[\left(\frac{\partial U^s_x}{\partial t}\right)^2 + \left(\frac{\partial U^s_y}{\partial t}\right)^2 + \left(\frac{\partial U^s_z}{\partial t}\right)^2\right] + \quad (20)$$

$$+ 2r^n_{22}\left[\left(\frac{\partial U^n_x}{\partial t}\right)^2 + \left(\frac{\partial U^n_y}{\partial t}\right)^2 + \left(\frac{\partial U^n_z}{\partial t}\right)^2\right]$$

The form (20) presupposes that the system is statistically isotropic. The mass coefficients $r_{11}, r^{s(n)}_{12}, r^{s(n)}_{22}$ take account of the additional motion which the solid and liquid acquire [22]. Therefore by analogy to Ref. 22 it is easily found that

$$f r^s = r^s_{22} + r^s_{12}, \quad f r^n = r^n_{22} + r^n_{12}, \quad (1-f) r_{sol} = r_{11} + r^s_{12} + r^n_{12} \quad (21)$$

Here

$$\mathbf{r}_{12}^{s} < 0, \quad \mathbf{r}_{12}^{n} < 0 \qquad (22)$$

are mass parameters characterising the increase in the inertial mass and, as usual, this increment to the inertial mass is called the associated mass. According to Berryman [34] the associated mass tensor is given by

$$\mathbf{r}_{12}^{s(n)} = -(\mathbf{a}_{\infty} - 1)\mathbf{f}\mathbf{r}^{s(n)} \qquad (23)$$

where $\alpha_{\infty} \geq 1$ is a geometric quantity, independent of the density of the solid and the density of the liquid, and can vary from 1 (for plane-parallel capillaries) to $\infty$ (for isolated pores or pores oriented perpendicular to the motion).

Let $q_x$ be the total force acting on the solid part of a unit volume and $Q_x^n$ and $Q_x^S$ the total forces acting on the superfluid and normal components per unit volume. Then the Lagrangian equations become

$$\frac{\partial}{\partial t}\left(\frac{\partial T}{\partial \dot{u}_x}\right) = \frac{\partial^2}{\partial t^2}\left(\mathbf{r}_{11} u_x + \mathbf{r}_{12}^{s} U_x^s + \mathbf{r}_{12}^{n} U_x^n\right) = q_x$$

$$\frac{\partial}{\partial t}\left(\frac{\partial T}{\partial \dot{U}_x^s}\right) = \frac{\partial^2}{\partial t^2}\left(\mathbf{r}_{12}^{s} u_x + \mathbf{r}_{22}^{s} U_x^s\right) = Q_x^s \qquad (24)$$

$$\frac{\partial}{\partial t}\left(\frac{\partial T}{\partial \dot{U}_x^n}\right) = \frac{\partial^2}{\partial t^2}\left(\mathbf{r}_{12}^{n} u_x + \mathbf{r}_{22}^{n} U_x^n\right) = Q_x^n$$

With no loss of generality we shall write the equations for the motion only along the $x$ direction. Expressing these forces in terms of the stress tensor

$$q_x = \frac{\partial \mathbf{s}_x}{\partial x} + \frac{\partial \mathbf{t}_z}{\partial y} + \frac{\partial \mathbf{t}_y}{\partial z}, \qquad Q_x^s = \frac{\partial s'}{\partial x}, \qquad Q_x^n = \frac{\partial s''}{\partial x} \qquad (25)$$

we obtain the dynamical equations

$$\frac{\partial s_x}{\partial x} + \frac{\partial t_z}{\partial y} + \frac{\partial t_y}{\partial z} = \frac{\partial^2}{\partial t^2}\left(r_{11} u_x + r_{12}^{s} U_x^{s} + r_{12}^{n} U_x^{n}\right)$$

$$\frac{\partial s'}{\partial x} = \frac{\partial^2}{\partial t^2}\left(r_{22}^{s} U_x^{s} + r_{12}^{s} u_x\right) \qquad (26)$$

$$\frac{\partial s''}{\partial x} = \frac{\partial^2}{\partial t^2}\left(r_{22}^{n} U_x^{n} + r_{12}^{n} u_x\right)$$

Similar equations can be obtained for motion along the $y$ and $z$ directions

In Eqs. (26) we neglected dissipative processes. To take such processes into account we shall assume that the main mechanism of dissipation in the system porous body-superfluid liquid is deceleration of the normal component of the superfluid liquid by the walls of the pores. Since all deformations are assumed to be small, the macroscopic motions studied in the theory are small elastic oscillations or waves. Consequently, as in most physical systems, the friction forces are proportional to the velocities of the moving physical points and can be described using a dissipative function. The dissipative function is, by definition, a homogeneous quadratic function of generalized velocities. The similarly to the Biot procedure, we obtain for the dissipative terms in the equations of motion of the solid and normal component of the superfluid liquid

$$N\nabla^2 \vec{u} + (A+N)\,grad\,e + Q^{s}\,grad\,e^{s} + Q^{n}\,grad\,e^{n} = \frac{\partial^2}{\partial t^2}\left(r_{11}\vec{u} + r_{12}^{s}\vec{U}^{s} + r_{12}^{n}\vec{U}^{n}\right) +$$

$$+ bF(w)\frac{\partial}{\partial t}\left(\vec{u} - \vec{U}^{n}\right)$$

$$Q^{s}\,grad\,e + R^{s}\,grad\,e^{s} + R^{sn}\,grad\,e^{n} = \frac{\partial^2}{\partial t^2}\left(r_{12}^{s}\vec{u} + r_{22}^{s}\vec{U}^{s}\right) \qquad (22)$$

$$Q^{n}\,grad\,e + R^{n}\,grad\,e^{n} + R^{sn}\,grad\,e^{s} = \frac{\partial^2}{\partial t^2}\left(r_{12}^{n}\vec{u} + r_{22}^{n}\vec{U}^{n}\right) - bF(w)\frac{\partial}{\partial t}\left(\vec{u} - \vec{U}^{n}\right)$$

The complex function $F(w)$ reflects the deviation from a Poiscuille flow taking account of the geometric features of the porous material [35], the coefficient $b = hf^{2}/k_0$ is the ratio of the total friction force to the average (relative to the solid) velocity of the normal component, $h$ is the viscosity of the fluid and $k_0$ is the permeability [22]. The possibilities of using the acoustics of a superfluid liquid to study various parameters of porous materials were analyzed theoretically in Ref. [35]. To this end the function $F(w)$ was expressed in terms of the key parameters: the tortuosity, the permeability, a dynamical parameter with the dimension of length, and the porosity. Some of them were obtained from the solution of the problem of the electric conductivity of a porous medium consisting of a insulating porous material filled with a conducting liquid. The response of the rigid porous medium was calculated. The results obtained make it possible to investigate the characteristic features of two-fluid hydrodynamics of $He\ II$ in a rigid porous medium and to determine the parameters of the medium from experimental data for the velocities of first, second, and fourth sounds.

Solids, in contrast to liquids, are characterized by, in addition to ordinary elasticity, elasticity with respect to shear deformations. Not only longitudinal but also transverse waves can exist even in an infinite solid medium [24]. The picture of wave propagation in our system is much richer than in a solid and in a superfluid liquid considered separately. Taking the curl of both sides of (27) equation gives:

$$N\nabla^2 \vec{rot\,u} = \frac{\partial^2}{\partial t^2}\left(r_{11}\vec{rot\,u} + r_{12}^S \vec{rot\,U^S} + r_{12}^n \vec{rot\,U^n}\right) + bF(w)\frac{\partial}{\partial t}\left(\vec{rot\,u} - \vec{rot\,U^n}\right)$$

$$r_{22}^S \vec{rot\,U^S} + r_{12}^S \vec{rot\,u} = 0 \qquad (28)$$

$$\frac{\partial^2}{\partial t^2}\left(r_{22}^n \vec{rot\,U^n} + r_{12}^n \vec{rot\,u}\right) - bF(w)\frac{\partial}{\partial t}\left(\vec{rot\,u} - \vec{rot\,U^n}\right) = 0$$

Following [12, 22] we shall write:

$$\vec{\Omega} = \vec{rot\,u}, \quad \vec{\Omega}^S = \vec{rot\,U^S}, \quad \vec{\Omega}^n = \vec{rot\,U^n} \qquad (29)$$

The following expression for the wave vector of transverse sound can be easily obtained from the system of equation (28):

$$q^2 = \frac{w^2}{N}\left(r_{sc} + \left(1 - a_\infty^{-1}\right)\Phi r^S + \Phi r^n \frac{(a_\infty - 1)\Phi r^n + ib\frac{F(w)}{w}}{a_\infty \Phi r^n + ib\frac{F(w)}{w}}\right). \qquad (30)$$

Here $r_{sc} = (1-f)r_{sol}$. Also for the vorticity we have:

$$\vec{\Omega}^S = \frac{a_\infty - 1}{a_\infty}\vec{\Omega}, \quad \vec{\Omega}^n = \frac{(a_\infty - 1)\Phi r^n + ib\frac{F(\vec{w})}{w}}{a_\infty \Phi r^n + ib\frac{F(w)}{w}}\vec{\Omega} \qquad (31)$$

McKenna et al. [9] modified the conventional two fluid hydrodynamic equations to take into account coupling of the normal component to the aerogel mass and elasticity. Their equations contain a vector restoring force and no shear stress tensor. They are therefore exact only for longitudinal waves in infinite media and differ to us is not allowed transverse wave solutions.

## 4. SOUND PROPAGATION IN UNRESTRICTED GEOMETRY

Now it will be interesting to ignore dissipative process in equation (27) and consider the case of unrestricted geometry. Then it is not difficult to obtain the equations of motion of elastic solid and hydrodynamic equations of superfluid helium from equation (27). For the first case we suppose $f = 0$ and ignore by the liquid part. Then we have

$$N\nabla^2 \vec{u} + (A+N)\,grad\,e = \frac{\partial^2}{\partial t^2} r_{sol}\vec{u} \qquad (32)$$

From this equation follows that elastic represents two independent propagating waves [24]. In one of them displacement is oriented along propagation of the wave, velocity of which is

$$C_l = \sqrt{\frac{A+2N}{r_{sol}}} \qquad (33)$$

and that is called longitudinal wave. Here $r_{sol}$ is density of solid. In another wave displacement is directed into the plane perpendicular to the direction of propagation, and is called as transverse wave, propagating with velocity

$$C_t = \sqrt{\frac{N}{r_{sol}}} \qquad (34)$$

If assuming that $a_\infty = f = 1$ and also ignoring by the part of solid equation, then from (27), we obtain the equation of superfluid motion

$$R^S \, grad \, e^S + R^{Sn} \, grad \, e^n = r^S \frac{\partial^2 \vec{U}^S}{\partial t^2}$$

$$R^n \, grad \, e^n + R^{Sn} \, grad \, e^S = r^n \frac{\partial^2 \vec{U}^n}{\partial t^2} \qquad (35)$$

In (35) general elastic coefficients have the form

$$R^S = \frac{r^S}{r}\left(r^S C_1^2 + r^n C_2^2\right)$$

$$R^n = \frac{r^n}{r}\left(r^n C_1^2 + r^S C_2^2\right) \qquad (36)$$

$$R^{Sn} = \frac{r^S r^n}{r}\left(C_1^2 - C_2^2\right)$$

So, for pure $HeII$ solution solving the system (35) in the usual manner we obtain the dispersion equation for the bulk waves propagating in free $HeII$ superfluid:

$$C^4 r^S r^n - C^2\left(r^S R^n + r^n R^S\right) + R^S R^n - \left(R^{Sn}\right)^2 = 0 \qquad (37)$$

Equations (37) has two roots:

$$C_1^2 = \frac{K_f}{r} \, , \qquad C_2^2 = \frac{r^S s^2 T}{r C_{He}} \qquad (38)$$

which conform to the velocity of the first and the second sounds correspondingly. From equation (35) follow the well known results for the fourth sound in superfluid $He^4$. If we assume $\vec{U}^n = 0$ in (35), we derive

$$C_4^2 = \frac{r^S}{r} C_1^2 + \frac{r^n}{r} C_2^2 \qquad (39)$$

Propagation of the fourth sound in a superfluid $He^4$ was studied in [3, 4] from the Landau hydrodynamic equations.

## 5. WAVE PROPAGATION IN THE HIGH-FREQUENCY LIMIT

Two important limiting cases are considered. The penetration depth of a viscous wave can be large or small compared with the pores sizes. The first case corresponds to low oscillatory frequencies of the thermodynamic quantities and the second case corresponds to high frequencies.

In the high-frequency limit the dissipative terms can be neglected, and the system of equations (27) can be used to investigate propagating waves.

The equations (28) show that the velocities of the solid and the superfluid liquid are intercoupled. In the high-frequency limit, when $|F(w)/w| \ll 1$, we have from (30):

$$C_t^2 = \frac{N}{r_{sc} + (1 - a_\infty^{-1}) f r} \tag{40}$$

Here $r_{sc} = (1-f) r_{sol}$. In the expression (40) the velocity becomes lower than for a «dry» material as a result of entrainment of a portion $(1 - a_\infty^{-1})$ of the liquid by the «framework» of the porous body because of the tortuosity and the associated increase in the effective density. In this wave the superfluid and normal components move as a single entity:

$$\vec{\Omega}^n = \vec{\Omega}^s = \frac{a_\infty - 1}{a_\infty} \vec{\Omega}, \tag{41}$$

and the liquid and solid rotate in the same direction.

Landau always supplemented the hydrodynamic equations of motion of a superfluid liquid by the requirement that the superfluid motion be potential: $rot \vec{V}^s = 0$. That is, the superfluid component cannot rotate as a rigid body. If in our theory the displacement (deformation vectors) $\vec{U}^s$ of the superfluid component occurred in a region greater in size than the size of atoms but much smaller than the distance between pores, then Landau's condition would be satisfied in this case. However, in the theory which we are proposing $\vec{U}^s$ is the displacement of the superfluid component averaged over an element much greater in size than the pore size. Consequently, there is no need for the motion to be irrotational. For this reason, the displacements $\vec{U}^s$ and $\vec{U}^n$ appear in the equations as equivalent quantities. The only difference between them is that the superfluid motion occurs without entropy transport. We note that in this work the vortex motion (41) of the superfluid liquid is not a complex rotation around vortex singularities, but rather it is due to the fact that the solid gives rise to partial entrainment of the liquid in the vortex motion.

To study longitudinal waves we shall consider all displacements as a gradient of some scalar function. Then, the following dispersion equation follows from the system of equations (27):

$$(a_\infty f r^s C^2 + \frac{r^s}{r^n} R^{sn} - R^s)\{[(r_{sc} + f r)a_\infty - f r] f r^n C^4 - r^n C^2[(\frac{r_{sc}}{r} + f)R - 2f(Q + R) +$$

$$+ a_\infty f(A + 2N + 2Q + R)] + R(A + 2N) - Q^2\} = 0 \qquad (42)$$

One solution of this dispersion equation is second sound (temperature wave) in a porous medium [8]:

$$C^2 = \frac{C_2^2}{\alpha_\infty}, \qquad (43)$$

where $C_2$ is the velocity of second sound in unbounded helium. In addition to this solution Eq. (42) possesses two other solutions which are similar to the fast and slow waves in a porous medium filled with an ordinary classical liquid.

For an unconsolidated porous medium ($K_b = N = 0$) the velocity of the slow wave, just as the velocity of the shear wave, is identically equal to zero even in the limit of high oscillation frequencies [31]:

$$C^2_{slow}(K_b = N = 0) = 0 \qquad (44)$$

Finally, the velocity of a fast compressible wave has the form [31]:

$$C^2_{fast}(K_b = N = 0) = \frac{a_\infty + f(\frac{r_{sc}}{r} + f - 2)}{[(r_{sc} + f r)a_\infty - f r][\frac{1-f}{K_s} + \frac{f}{K_f}]} \qquad (45)$$

For a rigid porous medium $K_f \ll K_b$, $N \ll K_s$. In this case, for the fast and slow modes we obtain from Eq. (42) the well-known expressions [31]:

$$C_{fast}^2 = \frac{K_b + \frac{4}{3}N}{r_{sc} + fr\left(1 - a_\infty^{-1}\right)} \tag{46}$$

$$C_{slow}^2 = \frac{C_1^2}{a_\infty} \tag{47}$$

where $C_1 = (\partial p/\partial r)^{1/2}$ is the volume propagation velocity of first sound in $He\,II$.

## 6. WAVE PROPAGATION IN THE LOW-FREQUENCY LIMIT

Since the penetration depth of the viscous wave becomes large in the limit of low oscillation frequencies, the normal component of the superfluid liquid completely sticks to the matrix of the porous medium, as a result of which the «framework» of the solid body and the normal liquid move with the same velocity $\vec{V}^n \mathbf{1}\, 0$. Obviously, to examine this situation the friction force must be eliminated in Eqs. (27) and $\vec{U}^n = \vec{u}$ must be set in the two equations obtained. This yield

$$N\nabla^2 \vec{U}^n + \left(A + N + 2\frac{r^n}{r}Q + R^n\right)graddiv\,\vec{U}^n + \left(\frac{r^S}{r}Q + R^{Sn}\right)graddiv\,\vec{U}^S =$$

$$= \frac{\partial^2}{\partial t^2}\left[(r_{11} + 2r_{12}^n + r_{22}^n)\vec{U}^n + r_{12}^S\vec{U}^S\right] \tag{48}$$

$$\left(\frac{r^S}{r}Q + R^{Sn}\right)graddiv\,\vec{U}^n + R^S graddiv\,\vec{U}^S = \frac{\partial^2}{\partial t^2}\left(r_{12}^S\vec{U}^n + r_{22}^S\vec{U}^S\right)$$

Just as in the preceding section, from Eq. (30) we can calculate the velocity of a transverse wave in the limit of low oscillation frequencies [32]:

$$C_t^2 = \frac{N}{(r_{sc} + fr^n) + (1 - a_\infty^{-1})fr^s} \tag{49}$$

For low oscillation frequencies the velocity of a transverse wave is lower than for the «dry» material by an even larger amount than at high frequencies. This is because the normal component of $He\,II$ now completely sticks to the «framework» of the porous body on account of viscosity, while the superfluid component is entrained only partially because of tortuosity, just as at high frequencies. The formula (49) makes it possible to determine the parameter $a_\infty$ directly. For the transverse wave from (31) we have:

$$\vec{\Omega}^s = \frac{a_\infty - 1}{a_\infty} \vec{\Omega}^n \tag{50}$$

For longitudinal waves we obtain the equation:

$$fr^s[a_\infty r_{sc} + f(a_\infty r - r^s)]C^4 - C^2[R^s(r_{sc} + fr) + fr^s a_\infty(A + 2N + 2Q + R) - 2fr^s \cdot (\frac{r^s}{r}Q + R^s + R^{sn})] + R^s(A + 2N + 2Q + R) - (\frac{r^s}{r}Q + R^s + R^{sn})^2 = 0 \tag{51}$$

The two solutions of this equation in the low-frequency limit are the velocities of fast and slow waves. So, the solution of the dispersion equation are only longitudinal waves in infinite media and do not allow transverse wave solutions. For an aerogel or, equivalently, for an open geometry $j \approx 1$, and $K_b << K_s$. Under these conditions, taking account of the fact that in a «framework» which is not filled with liquid the velocity of the longitudinal wave $C_a^2 = \frac{[K_b + (4/3)N]}{r_a}$, the dispersion equation (53) is identical to the dispersion equation obtained in Refs. [9].

The first solution is intermediate between the first and fourth sound

$$C_{14}^2 = \frac{C_1^2 + \frac{r^a}{r^n}C_4^2}{1 + \frac{r^a}{r^n}} \tag{52}$$

and it resembles the fast mode.

Another solution corresponds to the slow mode, which is an oscillation of a deformation of the aerogel combined with a simultaneous out-of-phase motion of the superfluid component:

$$C_{2a}^2 = \frac{C_2^2 + \frac{\rho^a \rho^S}{\rho \rho^n} C_a^2}{1 + \frac{\rho^a \rho^S}{\rho \rho^n}} \tag{53}$$

Second mode has been observed in $HeII$ in aerogel as heat pulse propagation with the velocity of first sound. It should be observed on impure superfluids-aerogel also. From experiment date for silica aerogel $C_a^2 \gg C_2^2$ [9], so from the above mentioned formula it follows that $C_{2a}^2 \gg C_2^2$ Therefore, the velocity of slow wave is much bigger than the velocity of temperature sound in $HeII$.

The theory developed in the paper makes it possible to find relations between quantities which oscillate in the waves. Specifically, the temperature and pressure oscillations are related with the displacement vector as

$$T' = \frac{\sigma \rho^S T}{\rho C_{He}} div\left(\vec{U}^S - \vec{U}^n\right) \tag{53}$$

$$p' = -\frac{K_f}{\rho} div\left(\rho^n \vec{U}^n + \rho^S \vec{U}^S\right) \tag{54}$$

Using these expressions, we obtain for aerogels at low frequencies:

$$\frac{p'}{T'} = \frac{K_f C_{He}}{\sigma T} \frac{\rho^n (C^2 - C_2^2)}{\rho^S (C^2 - C_1^2)}. \tag{55}$$

According to these formulas the pressure and temperature oscillate in fast and slow waves [12]. However, it should be noted that the main oscillatory quantity is the pressure in fast sound and the temperature in slow sound. Therefore, for identical temperature oscillations we have:

$$\frac{p'_{fast}}{p'_{slow}} = \frac{C_{fast}^2 - C_2^2}{C_{fast}^2 - C_1^2} \frac{C_{slow}^2 - C_1^2}{C_{slow}^2 - C_2^2} \approx \frac{C_1^2}{C_{slow}^2} \gg 1, \tag{56}$$

and for identical pressure oscillations in the waves $\left(T'_{slow}/p'_{slow}\right) \gg 1$ for the ratio of the values of the temperature. Likewise, the ratios $\left(p'/T'\right)_{slow}$ and $\left(T'/p'\right)_{fast}$ in aerogels are much larger than the corresponding ratios in pure helium. This shows that the temperature oscillations in first sound and pressure oscillations in second sound are much stronger in an aerogel filled with superfluid helium than in pure helium. It is well known that a similar phenomenon is observed in weak solutions of $He^3$ in $He^4$, this is due to the finite value of $\left(\partial r/\partial c\right)$ (c is $He^3$ concentration, $r$ - density of the solution). Namely, sound propagation in impure superfluids (superfluid $He^3-He^4$ solution can be considered as an example of an impure superfluid) has a number of peculiarities connected with the oscillations of pressure and temperature in the acoustic wave. Whereas in pure helium $II$ only the pressure oscillates in the first sound wave, and only the temperature oscillates in the second sound wave (neglecting the coefficient of thermal expansion, which is enormously small for helium), in a solution there are pressure, temperature, and concentration oscillations in both waves. In the first sound wave the oscillation of the temperature is proportional to the coefficient $b = \left(c/r\right)\left(\partial r/\partial c\right)$ and in the second sound wave the pressure oscillation is proportional to same coefficient. At low $He^3$ concentration the quantities proportional to $b$ can not be neglected. Unlike pure $He^4$, the first sound wave in solutions contains a relative oscillation of the normal and superfluid liquids, the magnitude of which is proportional to $b$. This results in peculiarities in method used for their excitation. This phenomenon was studied first, theoretically and experimentally, in Ref. 36. The similar phenomena is observed in system aerogel-$HeII$. In this system the coupling between oscillating parameters is provided by $r^a/r$. The general case superfluid $He^3 - He^4$ solution-porous media was considered in [37].

One solution for a rigid body, corresponding to a wave propagating in the «framework» is modified because the normal component sticks to the «framework» and because a nonviscous superfluid component is present. It is expressed as follows:

$$C_{fast}^2 = \frac{K_b + \frac{4}{3}N}{r_{sc} + fr - f a_\infty^{-1} r^s} \quad (58)$$

The second solution is fourth sound in a porous medium [38]

$$C_{slow}^2 = \frac{C_4^2}{a_\infty} \quad (59)$$

In Ref. [38] a correction factor n, equal to the ratio of the true to the measured value of the velocity of fourth sound, was introduced to determine the velocity of fourth sound and a filter of highly compressed $Fe_2O_3$ powder was used to observe the sound..

In summary, we have obtained the hydrodynamic equations for the three-component system porous body-superfluid liquid $He^4$, using which enabled us to analyze the propagation of longitudinal and transverse waves in the low and high-frequency limits for any values of the porosity and elasticity coefficients. The elasticity coefficients appearing in the equations were expressed in terms of physically measurable quantities. The derived equations were applied to the most important particular case when the normal component of superfluid helium is locked inside a highly porous media (aerogel, Im-helium sample) by viscous forces. For this case the velocities of two longitudinal sound modes and transverse mode were calculated from the derived equations. There were established the coupling between temperature and pressure oscillations in these fast and slow modes.

# L I T E R A T U R E